\documentclass[12pt,letterpaper]{article}
\usepackage{amsmath,amsfonts,amsthm,amssymb}

\theoremstyle{plain}

\numberwithin{equation}{section}
\newtheorem{thm}{Theorem}[section]
\newtheorem{lem}[thm]{Lemma}
\newtheorem{cor}[thm]{Corollary}

\newenvironment{exam}[1]%
{\begin{flushleft}\textbf{Example #1}.\enspace}%
{\end{flushleft}}

\newcounter{cond}

\newcommand{\complex}{{\mathbb C}}
\newcommand{\real}{{\mathbb R}}
\newcommand{\ascript}{{\mathcal A}}
\newcommand{\bscript}{{\mathcal B}}
\newcommand{\pscript}{{\mathcal P}}

\newcommand{\overnub}{\,{\overline{\nu(B)}}\,}

\newcommand{\cupdot}{\mathbin{\cup{\hskip-5.4pt}^\centerdot}\,}
\newcommand{\subcupdot}{\mathbin{\cup{\hskip-4pt}^\centerdot}\,}
\newcommand{\bigcupdotin}{{\bigcup _{i=1}^n}{\hskip-8pt}^\centerdot{\hskip 8pt}}
\newcommand{\bigcupdotitwon}{{\bigcup _{i=2}^n}{\hskip-8pt}^\centerdot{\hskip 8pt}}
\newcommand{\bigcupdotif}{{\bigcup _{i=1}^\infty}{\hskip-8pt}^\centerdot{\hskip 8pt}}

\newcommand{\rmtr}{\mathrm{tr}}
\newcommand{\rmre}{\mathrm{Re}}

\newcommand{\ab}[1]{\left|#1\right|}
\newcommand{\brac}[1]{\left\{#1\right\}}
\newcommand{\paren}[1]{\left(#1\right)}
\newcommand{\sqbrac}[1]{\left[#1\right]}
\newcommand{\elbows}[1]{{\left\langle#1\right\rangle}}
\newcommand{\sqpar}[1]{\left[#1\right)}

\begin{document}

\title{QUANTUM MEASURE\\ and INTEGRATION THEORY}
\author{Stan Gudder\\ Department of Mathematics\\
University of Denver\\ Denver, Colorado 80208\\
sgudder@math.du.edu}
\date{}
\maketitle

\begin{abstract}
This article begins with a review of quantum measure spaces. Quantum forms and indefinite inner-product spaces are then discussed. The main part of the paper introduces a quantum integral and derives some of its properties. The quantum integral's form for simple functions is characterized and it is shown that the quantum integral generalizes the Lebesgue integral. A bounded, monotone convergence theorem for quantum integrals is obtained and it is shown that a Radon-Nikodym type theorem does not hold for quantum measures. As an example, a quantum-Lebesgue integral on the real line is considered.
\end{abstract}

\section{Introduction}  
Quantum measure theory was introduced by R.~Sorkin in his studies of the histories approach to quantum mechanics and quantum gravitation \cite{sor94, sor07}. Since its inception in 1994, other researchers have also contributed to the field \cite{sal02, scmslsw08, sw08}. Sorkin's work and subsequent investigations only considered finite quantum measure spaces until the author studied the general theory of quantum measure spaces \cite{gud1}. Not only are there applications of this field to the study of decoherence functionals and the histories approach to quantum mechanics, but the author has suggested that it might be applied to the computation and prediction of elementary particle masses \cite{gud1}.

This article begins with a review of quantum measure spaces and summarizes some of the results in
\cite{gud1, gud2}. We then discuss quantum forms and indefinite inner-product spaces. We next proceed to the main part of the paper which introduces a quantum integral and derives some of its properties. In particular, we characterize the quantum integral's form for simple functions and show that the quantum integral generalizes the Lebesgue integral in the sense that if the quantum measure is an ordinary measure, then the quantum integral reduces to the Lebesgue integral. We obtain a bounded, monotone convergence theorem for quantum integrals and show that a Radon-Nikodym type theorem does not hold for quantum measures. As an example, a quantum Lebesgue integral on the real line is considered and a quantum fundamental theorem of calculus for this integral is proved. We mention in passing that generalizations to super-quantum measures and integrals are possible but leave these for future investigations. Although many of our results extend to quantum measures that can have a value $+\infty$, for technical reasons we restrict attention to finite quantum measures.

\section{Quantum Measure Spaces} 
A sequence of sets $A_i$ is \textit{increasing} if $A_i\subseteq A_{i+1}$ and \textit{decreasing} if
$A_i\supseteq A_{i+1}$, $i=1,2,\ldots\,$. If two sets $A$ and $B$ are disjoint we use the notation $A\cupdot B$ for
$A\cup B$. Recall that a \textit{measurable space} is a pair $(X,\ascript )$ where $X$ is a nonempty set and $\ascript$ is a $\sigma$-algebra of subsets of $X$. A (finite) \textit{measure} on $\ascript$ is a nonnegative set function
$\mu\colon\ascript\to\real ^+$ that satisfies
\begin{list} {(\arabic{cond})}{%
\setlength\itemindent{-7pt}}
\item [{(1)}]   
$\mu (A\cupdot B)=\mu (A)+\mu (B)$ for any disjoint $A,B\in\ascript$ (\textit{additivity})
\item [{(2)}]   
if $A_i\in\ascript$ is an increasing sequence, then
\begin{equation*}
\mu (\cup A_i)=\lim\mu (A_i)\quad\hbox{(\textit{continuity})}
\end{equation*}
\end{list}
For reasons that will soon be clear, we also call (1) \textit{grade}-1 \textit{additivity}. Conditions (1) and (2) together are equivalent to $\sigma$-\textit{additivity}:
\begin{equation*}
\mu\paren{\bigcupdotif A_i}=\sum _{i=1}^\infty\mu (A_i)
\end{equation*}
Moreover, it follows (1) and (2) that
\begin{list} {(\arabic{cond})}{%
\setlength\itemindent{-7pt}}
\item [{(3)}]   
if $A_i\in\ascript$ is a decreasing sequence, then
\begin{equation*}
\mu (\bigcap A_i)=\lim\mu (A_i)
\end{equation*}
\end{list}

Because of quantum interference, the additivity condition (1) does not hold for quantum measures. Instead we have the weaker condition
\begin{list} {(\arabic{cond})}{%
\setlength\itemindent{-7pt}}
\item [{(4)}]   
$\mu (A\cupdot B\cupdot C)=\mu (A\cupdot B)+\mu (A\cupdot C)+\mu (B\cupdot C)
   \!-\!\mu (A)\!-\!\mu (B)\!-\!\mu (C)$\break
  {\hglue -8pt}for all mutually disjoint $A,B,C\in\ascript$.
\end{list}
We call (4) \textit{grade}-2 \textit{additivity}. A grade-2 additive set function $\mu\colon\ascript\to\real ^+$ that satisfies (2) and (3) is called a $q$-\textit{measure}. If $\mu$ is a $q$-measure on $\ascript$, then $(X,\ascript ,\mu )$ is a
$q$-\textit{measure space}.

One can also consider super-quantum measures called \textit{grade}-$n$ \textit{measures}, $n=2,3,\ldots\,$. These satisfy (2), (3) and the grade-$n$ additivity condition:
\begin{align*}
\mu (A_1\cupdot\cdots\cupdot A_{n+1})
&=\sum _{i_1<\cdots <i_n=1}^{n+1}\mu (A_{i_1}\cupdot\cdots\cupdot A_{i_n})\\
&\quad -\sum _{i_1<\cdots <i_{n-1}=1}\mu (A_{i_1}\cupdot\cdots\cupdot A_{I_{n-1}})\\
&\quad +\cdots +(-1)^{n+1}\sum _{i=1}^n\mu (A_i)
\end{align*}
It can be shown by induction that a grade-$n$ measure is a grade-$(n+1)$ measure, $n=1,2,\ldots$, and we then obtain a hierarchy of types of measures. Although much of our work generalizes to super-quantum measures, we shall only consider $q$-measures here.

A simple example of a $q$-measure is the square of a measure. Thus, if $\nu$ is a measure on $\ascript$, then $\mu$ defined by $\mu (A)=\nu (A)^2$ for all $A\in\ascript$ is a $q$-measure. A slightly more general example is
$\mu (A)=\ab{\nu (A)}^2$ where $\nu$ is a complex-valued measure on $\ascript$. This last example is applicable to quantum mechanics when $\nu$ describes a quantum amplitude measure. In this case,
\begin{align*}
\mu (A\cupdot B)&=\ab{\nu (A\cupdot B)}^2=\ab{\nu (A)+\nu (B)}^2\\
&=\mu (A)+\mu (B)+2\rmre\sqbrac{\nu (A)\overnub}
\end{align*}
Additivity is destroyed by the quantum interference term $2\rmre\sqbrac{\nu (A)\overnub}$.

A $q$-measure $\mu$ is \textit{regular} if
\begin{list} {(\arabic{cond})}{%
\setlength\itemindent{-7pt}}
\item [{(5)}]   
$\mu (A)=0$ implies that $\mu (A\cupdot B)=\mu (B)$
\item [{(6)}]   
$\mu (A\cupdot B)=0$ implies $\mu (A)=\mu (B)$
\end{list}
If $\mu$ is regular and $\mu (A)=0$ implies $\mu (B)=0$ for all $B\in\ascript$ with $B\subseteq A$, then $\mu$ is
\textit{completely regular}. It is clear that our previous two examples of $q$-measures are completely regular.

We now discuss a considerably more general example of a $q$-measure. This example comes from the concept of a decoherence functional which is the original motivation for the study of $q$-measures \cite{sor94, sor07, sw08}.
A \textit{decoherence functional} is a map $D\colon\ascript\times\ascript\to\complex$ that satisfies
\begin{list} {(\arabic{cond})}{%
\setlength\itemindent{-7pt}}
\item [{(7)}]   
$D(A\cupdot B,C)=D(A,C)+D(B,C)$
\item [{(8)}]   
$D(A,B)=\,\overline{D(B,A)}\,$
\item [{(9)}]   
$D(A,A)\ge 0$
\end{list}
\begin{list} {(\arabic{cond})}{%
\setlength\itemindent{-1pt}}
\item [{(10)}]   
$\ab{D(A,B)}^2\le D(A,A)D(B,B)$.
\end{list}
Thus, a decoherence functional is like an inner-product on sets. We say that $D$ is \textit{continuous} if
$\mu (A)=D(A,A)$ satisfies (2) and (3). An example of a continuous decoherence functional that comes up in quantum measurement theory is $D(A,B)=\rmtr\sqbrac{WE(A)E(B)}$ where $W$ is a density operator (state) and $E$ is a positive operator-valued measure (observable). It can be shown that for any continuous decoherence functional $D$ we have that $\mu (A)=D(A,A)$ is a completely regular $q$-measure \cite{gud1, gud2, sw08}.

Besides these theoretical reasons, there are also experimental reasons for considering $q$-measures. In the
well-known two-slit experiment, a beam of particles is directed toward a screen containing two slits $A_1$ and $A_2$ and the particles that pass through the slits impinge upon a detection screen $S$. If $\mu (A_i)$ denotes the probability that a particle hits a small region $\Delta\subseteq S$ after passing through slit $A_i$, $i=1,2$, then $\mu (A_1\cupdot A_2)\ne\mu (A_1)+\mu (A_2)$ in general. Thus, $\mu$ is not additive. However, recent experiments involving a three slit screen indicate that $\mu$ is grade-2 additive \cite{scmslsw08}. Hence, $\mu$ is a $q$-measure.

We now give some simple examples of $q$-measure spaces. We use the notation $\mu (x)=\mu\paren{\brac{x}}$ for singleton sets $\brac{x}$.
\begin{exam}{1}    
Let $X=\brac{x_1,x_2,x_3,x_4}$ and let $\pscript (X)$ be the power set on $X$. Define the measure $\nu$ on
$\pscript (X)$ by $\nu (x_i)=1/2$, $i=1,2,3,4$. We may think of $X$ as the four outcomes of flipping a fair coin twice. Then $\nu (A)$ is the probability that event $A\in\pscript (X)$ occurs. For example, if $A=\brac{x_1,x_2,x_3}$ is the event that at least one head appears in the two flips, then $\nu (A)=3/4$. Now we consider the ``quantum coin'' with ``probabilities'' given by $\mu (A)=\nu (A)^2$. In this case the ``probability'' of each sample point $x_i$ is $1/16$ and
$\mu (A)=9/16$. As mentioned earlier, $\mu$ is a completely regular $q$-measure.
\end{exam}

\begin{exam}{2}    
Let $X=\brac{x_1,x_2,x_3}$ with $\mu (\emptyset )=\mu (x_1)=0$ and $\mu (A)=1$ for all other $A\in\pscript (X)$. Then $\mu$ is a completely regular $q$-measure on $\pscript (X)$.
\end{exam}

\begin{exam}{3}    
Let $X=\brac{x_1,\ldots ,x_m,y_1,\ldots ,y_m,z_1,\ldots ,z_n}$ and call $(x_i,y_i)$, $i=1,\ldots ,m$,
\textit{destructive pairs} (or \textit{particle-antiparticle pairs}). Denoting the cardinality of a set $B$ by $\ab{B}$ we define
\begin{equation*}
\mu (A)=\ab{A}-2\ab{\brac{(x_i,y_i)\colon x_i,y_i\in A}}
\end{equation*}
for every $A\in\pscript (X)$. Thus, the $\mu$ measure of $A$ is the cardinality of $A$ after the destructive pairs of $A$ annihilate each other. For instance, $\mu\paren{\brac{x_1,y_1,z_1}}=1$, $\mu\paren{\brac{x_1,y_1,y_2,z_1}}=2$. Then $\mu$ is a regular but not completely regular $q$-measure on $\pscript (X)$.
\end{exam}

\begin{exam}{4}    
This is a continuum generalization of Example~3. Let $X=\sqbrac{0,1}$, let $\nu$ be Lebesgue measure on $X$ and let $\bscript (X)$ be the $\sigma$-algebra of Borel subsets of $X$. Define $\mu\colon\bscript (X)\to\real ^+$ by
\begin{equation*}
\mu (A)=\nu (A)-2\nu\paren{\brac{x\in A\colon x+3/4\in A}}
\end{equation*}
In this case, pairs $(x_,x+3/4)$ with $x\in A$ and $x+3/4\in A$ act as destructive pairs. For instance, $\mu (X)=1/2$,
$\mu \paren{\sqbrac{0,1/4}\cupdot\sqbrac{3/4,1}}=0$ and $\mu\paren{\sqbrac{0,1/4}}=1/4$. Again, $\mu$ is regular but not a completely regular $q$-measure on $\bscript (X)$.
\end{exam}

We now summarize some of the known results concerning $q$-measure spaces. The \textit{symmetric difference} of sets $A$ and $B$ is $A\Delta B=(A\cap B')\cupdot (A'\cap B)$ where $A'$ denotes the complement of the set $A$. Notice that $A\Delta B=A\cup B\smallsetminus A\cap B$. The next result is the quantum counterpart to the usual formula
\begin{equation*}
\nu (A\cup B)=\nu (A)+\nu (B)-\nu (A\cap B)
\end{equation*}
for measures \cite{gud1, gud2}.

\begin{thm}       
\label{thm21}
A map $\mu\colon\ascript\to\real ^+$ is grade-2 additive if and only if $\mu$ satisfies
\begin{equation*}
\mu (A\cup B)=\mu (A)+\mu (B)-\mu (A\cap B)+\mu (A\Delta B)-\mu (A\cap B')-\mu (A'\cap B)
\end{equation*}
\end{thm}

The following shows that grade-2 additivity can be extended to more than three mutually disjoint sets
\cite{gud1,gud2,sal02}.

\begin{thm}       
\label{thm22}
If $\mu\colon\ascript\to\real ^+$ is grade-2 additive, then for any $n\ge 3$ we have
\begin{equation*}
\mu\paren{\bigcupdotin A_i}=\sum _{i<j=1}^n\mu (A_i\cupdot A_j)-(n-2)\sum _{i=1}^n\mu (A_i)
\end{equation*}
\end{thm}

Let $(X,\ascript ,\mu )$ be a $q$-measure space. We say that $A,B\in\ascript$ are $\mu$-\textit{compatible} and write $A\mu B$ if
\begin{equation*}
\mu (A\cup B)=\mu (A)+\mu (B)-\mu (A\cap B)
\end{equation*}
If $A\mu B$ then $\mu$ acts like a measure on $A\cup B$ so in some weak sense, $A$ and $B$ do not interfere with each other. Clearly, $A\mu A$ for every $A\in\ascript$. It follows from Theorem~\ref{thm21} that $A\mu B$ if and only if
\begin{equation*}
\mu (A\Delta B)=\mu (A\cap B')+\mu (A'\cap B)
\end{equation*}
The $\mu$-\textit{center} of $\ascript$ is
\begin{equation*}
Z_\mu =\brac{A\in\ascript\colon A\mu B\hbox{ for all }B\in\ascript}
\end{equation*}
A set $A\in\ascript$ is $\mu$-\textit{splitting} if $\mu (B)=\mu (B\cap A)+\mu (B\cap A')$ for all $B\in\ascript$
\cite{gud1,gud2}.

\begin{thm}       
\label{thm23}
{\rm (a)}\enspace $A$ is $\mu$-splitting if and only if $A\in Z_\mu$.
{\rm (b)}\enspace $Z_\mu$ is a sub $\sigma$-algebra of $\ascript$ and the restriction $\mu\mid Z_\mu$ is a measure. Moreover, if $A_i\in Z_\mu$, $i=1,2,\ldots$, are mutually disjoint, then for every $B\in\ascript$ we have
\begin{equation*}
\mu\sqbrac{\cupdot (B\cap A_i)}=\sum _\mu (B\cap A_i)
\end{equation*}
\end{thm}

For Examples~3 and ~4 we have the following result.

\begin{thm}       
\label{thm24}
The following statements are equivalent:
{\rm (a)}\enspace $A\in Z_\mu$,\newline
{\rm (b)}\enspace $A\mu A'$,
{\rm (c)}\enspace $\mu (A)+\mu (A')=1/2$.
\end{thm}

If $(X,\ascript )$ is a measurable space, we can form the Cartesian product $(X\times X,\ascript\times\ascript )$ which becomes a measurable space by letting $\ascript\times\ascript$ be the $\sigma$-algebra generated by the product sets $A\times B$, $A,B\in\ascript$. A signed measure $\lambda$ on $\ascript\times\ascript$ is \textit{symmetric} if
$\lambda (A\times B)=\lambda (B\times A)$ for all $A,B\in\ascript$. Also $\lambda$ is \textit{diagonally positive} if
$\lambda (A\times A)\ge 0$ for all $A\in\ascript$. The next result characterizes $q$-measures \cite{gud1}.

\begin{thm}       
\label{thm25}
A set function $\mu\colon\ascript\to\real ^+$ is a $q$-measure if and only if there exists a diagonally positive, symmetric signed measure $\lambda$ on $\ascript\times\ascript$ such that $\mu (A)=\lambda (A\times A)$ for all
$A\in\ascript$. Moreover, $\lambda$ is unique.
\end{thm}

\section{Quadratic and Quantum Forms} 
According to Theorem~\ref{thm25}, a $q$-measure is the diagonal part of a diagonally positive, symmetric signed measure. This indicates that a quantum integral should be a quadratic form which has even been suggested by R.~Sorkin \cite{sor07}. Also, the distinguishing feature of a $q$-measure is grade-2 additivity. We now show that a quadratic form has an analogous property.

In this section, $V$ will denote a real topological linear space. Although our work generalizes to complex spaces for simplicity we shall only consider the real case. A \textit{symmetric bilinear form} on $V$ is a map
$B\colon V\times V\to\real$ that satisfies
\begin{list} {(B\arabic{cond})}{\usecounter{cond}
\setlength{\rightmargin}{\leftmargin}}
\item $B(u,v)=B(v,u)$
\item $B(\alpha u,v)=\alpha B(u,v)$ for all $\alpha\in\real$
\item $B(u+v,w)=B(u,w)+B(v,w)$
\end{list}
The \textit{associated quadratic form} for $B$ is the map $Q\colon V\to\real$ given by $Q(v)=B(v,v)$. It is clear that
$Q (\alpha v)=\alpha ^2Q(v)$ for all $\alpha\in\real$ and if $B$ is continuous, then so is $Q$. In particular, $Q$ is even in the sense that $Q(-v)=Q(v)$.

\begin{lem}       
\label{lem31}
If $Q$ is the associated quadratic form for $B$, then
\begin{equation*}
B(u,v)=\tfrac{1}{2}\sqbrac{Q(u+v)-Q(u)-Q(v)}=\tfrac{1}{4}\sqbrac{Q(u+v)-Q(u-v)}
\end{equation*}
\end{lem}
\begin{proof}
For any $u,v\in V$ we have
\begin{align*}
Q(u+v)-Q(u)-Q(v)&=B(u+v,u+v)-B(u,u)-B(v,v)\\
   &=2B(u,v)
\end{align*}
Similarly,
\begin{align*}
Q(u+v)-Q(u-v)&=B(u+v,u+v)-B(u-v,u-v)\\
   &=4B(u,v)\qedhere
\end{align*}
\end{proof}

\begin{cor}       
\label{cor32}
Let $B_1,B_2$ be symmetric bilinear forms with associated quad\-ratic forms $Q_1$, $Q_2$, respectively. If $Q_1=Q_2$, then $B_1=B_2$.
\end{cor}

\begin{lem}       
\label{lem33}
If $Q$ is an associated quadratic form for $B$, then $Q$ satisfies:
{\rm (a)}\enspace $Q(u+v)+Q(u-v)=2\sqbrac{Q(u)+Q(v)}$
{\rm (b)}\enspace $Q(u+v+w)=Q(u+v)+Q(u+w)+Q(v+w)-Q(u)-Q(v)-Q(w)$
\end{lem}
\begin{proof}
To prove (a) we have
\begin{align*}
Q(u+v)&+Q(u-v)\\
   &=B(u+v,u+v)+B(u-v,u-v)\\
   &=B(u,u)+B(v,v)+2B(u,v)+B(u,u)+B(v,v)-2B(u,v)\\
   &=2\sqbrac{Q(u)+Q(v)}
\end{align*}
To prove (b) we have
\begin{align*}
Q(u+v)+&Q(u+w)+Q(v+w)-Q(u)-Q(v)-Q(w)\\
  &=B(u+v,u+v)+B(u+w,u+w)+Q(v+w,v+w)\\
  &\quad -B(u,u)-B(v,v)-B(w,w)\\
  &=B(u,u)+B(v,v)+B(w,w)+2\sqbrac{B(u,u)+B(v,v)+B(w,w)}\\
  &=B(u+v+w,u+v+w)=Q(u+v+w)\qedhere
\end{align*}
\end{proof}

Condition (a) in Lemma~\ref{lem33} is the well-known \textit{parallelogram law} and for obvious reasons, we call Condition~(b) in Lemma~\ref{lem33} \textit{grade}-2 \textit{additivity}. For example, if $B(u,v)=\elbows{u,v}$ is the inner-product for a real inner-product space then $B$ is a symmetric bilinear form with associated quadratic form given by the norm squared $Q(v)=\|v\|^2$. Now $Q$ is not additive in the sense that $\|u+v\|^2\ne\|u\|^2+\|v\|^2$ in general. However, by Lemma~\ref{lem33}(b), $Q$ is grade-2 additive.

\begin{thm}       
\label{thm34}
If $Q\colon V\to\real$ is continuous, then the following statements are equivalent.
{\rm (a)}\enspace $Q$ is an associated quadratic form.
{\rm (b)}\enspace $Q$ satisfies the parallelogram law.
{\rm (c)}\enspace $Q$ is even and grade-2 additive.
\end{thm}
\begin{proof}
By Lemma~\ref{lem33}, (a) implies both (b) and (c). To prove that (b) implies (a) define
$B\colon V\times V\to\complex$ by
\begin{equation*}
B(u,v)=\tfrac{1}{2}\sqbrac{Q(u+v)-Q(u)-Q(v)}
\end{equation*}
By the parallelogram law with $u=v=0$ we have $2Q(0)=4Q(0)$ so $Q(0)=0$. Again, by the parallelogram law with $u=v$ we have $Q(2u)=4Q(u)$. Hence,
\begin{equation*}
Q(u)=\tfrac{1}{2}\sqbrac{Q(2u)-2Q(u)}=B(u,u)
\end{equation*}
Thus, it suffices to show that $B$ is bilinear since it is clear that $B$ is symmetric. To show that $B$ is additive, we have by the parallelogram law that
\begin{align}
\label{eq31}                 
B(x+x',y)+B(x-x',y)&=\tfrac{1}{2}\sqbrac{Q(x+x'+y)-Q(x+x')-Q(y)}\notag\\
   &\quad +\tfrac{1}{2}\sqbrac{Q(x-x'+y)-Q(x-x')-Q(y)}\notag\\
   &=Q(x+y)+Q(x')-Q(x)-Q(x')-Q(y)\notag\\
   &=2B(x,y)
\end{align}
Letting $x'=x$ in \eqref{eq31} gives $B(2x,y)=2B(x,y)$. For $u,v\in V$, let $x=\frac{1}{2}(u+v)$, $x'=\frac{1}{2}(u-v)$ in \eqref{eq31}. Since $B(x/2,y)=B(x,y)/2$ we obtain
\begin{equation}
\label{eq32}                 
B(u,y)+B(v,y)=2B\paren{\tfrac{1}{2}(u+v),y}=B(u+v,y)
\end{equation}
Hence, $B$ is additive. By a standard argument employing \eqref{eq32} we have that $B(\alpha u,v)=\alpha B(u,v)$ for any positive rational number $\alpha$. By continuity this equation holds for any positive real number. Finally, letting  $v=-u$ in \eqref{eq32} gives $B(-u,y)=-B(u,y)$ so the equation holds for any negative real number. We next show that (c) implies (b). Letting $w=-v$ in the grade-2 additivity condition (Lemma~\ref{lem33}(b)) we obtain
\begin{equation*}
Q(u)=Q(u+v)+Q(u-v)-Q(u)-Q(v)-Q(-v)
\end{equation*}
Since $Q(-v)=Q(v)$ we obtain the parallelogram law.
\end{proof}

\begin{thm}       
\label{thm35}
If $Q\colon V\to\real$ is grade-2 additive, then for $n\ge 3$ we have
\begin{equation*}
Q\paren{\sum _{i=1}^nv_i}=\sum _{i<j=1}^nQ(v_i+v_j)-(n-2)\sum _{i=1}^nQ(v_i)
\end{equation*}
\end{thm}
\begin{proof}
We prove the result by induction on $n$. The result holds for $n=3$ by definition. Assuming the result holds for $n-1\ge 3$ we have
\begin{align*}
Q\paren{\sum _{i=1}^nv_i}&=Q\sqbrac{v_1+\cdots v_{n-2}+(v_{n-1}+v_n)}\\
  &=\sum _{i<j=1}^{n-2}Q(v_i+v_j)+\sum _{i=1}^{n-2}Q\sqbrac{v_i+(v_{n-1}+v_n)}\\
  &\quad -(n-3)\sqbrac{\sum _{i=1}^{n-2}Q(v_i)+Q(v_{n-1}+v_n)}\\
  &=\sum _{i<j=1}^{n-2}Q(v_i+v_j)+\sum _{i=1}^{n-2}Q(v_i+v_{n-1})+\sum _{i=1}^{n-2}Q(v_i+v_n)\\
  &\quad +(n-2)Q(v_{n-1}+v_n)-\sum _{n=1}^{n-2}Q(v_i)-(n-2)Q(v_{n-1})\\
  &\quad -(n-2)Q(v_n)-(n-3)\sqbrac{\sum _{i=1}^{n-2}Q(v_i)+Q(v_{n-1}+v_n)}\\
  &=\sum _{i<j=1}^nQ(v_i+v_j)-(n-2)\sum _{i=1}^nQ(v_i)
\end{align*}
The result follows by induction.
\end{proof}

By applying Theorem~\ref{thm35} or by employing mathematical induction, one can show that if $Q$ is grade-2 additive, then $Q$ is \textit{grade}-$n$ \textit{additive} for $n\ge 2$ in the sense that
\begin{align*}
Q\paren{\sum _{i=1}^{n+1}v_i}&=\sum _{i_1<\cdots <i_n=1}^{n+1}Q(v_{i_1}+\cdots +v_{i_n})
  -\sum _{i_1<\cdots <i_n=1}^{n+1}Q(v_{i_1}+\cdots +v_{i_{n-1}})\\
  &\quad +\cdots +(-1)^{n+1}\sum _{i=1}^{n+1}Q(v_i)
\end{align*}
Instead of giving the general proof we shall show that $Q$ satisfies grade-3 additivity. Indeed, from
Theorem~\ref{thm35} we have
\begin{align*}
\sum _{i<j<k=1}^4&Q(v_i+v_j+v_k)-\sum _{i<j=1}^4Q(v_i+v_j)+\sum _{i=1}^4Q(v_i)\\
  &=2\sum _{i<j=1}^4Q(v_i+v_j)-3\sum _{i=1}^4Q(v_i)-\sum _{i<j=1}^4Q(v_i+v_j)+\sum _{i=1}^4Q(v_i)\\
  &=\sum _{i<j=1}^4Q(v_i+v_j)-2\sum _{i=1}^4Q(v_i)=Q(v_1+v_2+v_3+v_4)
\end{align*}

Now let $(X,\ascript,\mu )$ be a $q$-measure space. By Theorem~\ref{thm25} there exists a diagonally positive, symmetric signed measure $\nu$ on $\ascript\times\ascript$ such that $\mu (A)=\nu (A\times A)$ for all $a\in\ascript$. Let $L_2(X^2)$ be the set of measurable functions $f\colon X^2\to\real$ such that $\int\sqbrac{f(x,y)}^2d\nu (x,y)$ exists and is finite. Although $L_2(X^2)$ is not a Hilbert space in general, we can define an indefinite inner product
\begin{equation*}
B(f,g)=\elbows{f,g}=\int fgd\nu
\end{equation*}
on $L_2(X^2)$. Then $L_2(X^2)$ becomes a topological linear space in the usual way and $B$ is a continuous, symmetric bilinear form on $L_2(X_2)$. The associated quadratic form is given by
\begin{equation*}
Q(f)=B(f,f)=\int f^2d\nu
\end{equation*}
Of course, $Q$ need not be nonnegative. It follows from Corollary~\ref{cor32} and Lemma~\ref{lem33} that $B$ is determined by $Q$ and that $Q$ is grade-2 additive.

The signed measure $\nu$ induces a \textit{marginal} signed measure
\begin{equation*}
\nu _1(A)=\nu (A\times X)=\nu (X\times A)
\end{equation*}
on $\ascript$. Let $L_2(X)$ be the set of measurable functions $f\colon X\to\real$ such that
\begin{equation*}
\int\sqbrac{f(x)}^2d\nu (x,y)=\int\sqbrac{f(x)}^2d\nu _1(x)
\end{equation*}
exists and is finite. Of course, $L_2(X)$ is a linear subspace of $L_2(X^2)$. We define the \textit{quantum form}
$Q_q\colon L_2(X)\to\real$ by
\begin{equation*}
Q_q(f)=\int f(x)f(y)d\nu (x,y)
\end{equation*}
Then $Q_q$ is a quadratic form associated to the symmetric bilinear form
\begin{equation*}
B_1(f,g)=\int f(x)g(y)d\nu (x,y)
\end{equation*}
so by Lemma~\ref{lem33}, $Q_q$ is grade-2 additive. Moreover, $Q_q$ generalizes the $q$-measure $\mu$ in the following sense. If $\chi _A$ is the characteristic function for $A\in\ascript$, then
\begin{equation*}
Q_q (\chi _A)=\int\chi _A(x)\chi _A(y)d\nu (x,y)=\nu (A\times A)=\mu (A)
\end{equation*}

As an example, let $\nu _1$ be a signed measure on $\ascript$ and form the product signed measure
$\nu =\nu _1\times\nu _1$ on $\ascript\times\ascript$. Thus, $\nu (A\times B)=\nu _1(A)\nu _1(B)$ for all
$A,B\in\ascript$ so $\nu$ is a diagonally positive, symmetric signed measure on $\ascript\times\ascript$. Corresponding to the $q$-measure $\mu (A)=\nu (A\times A)=\nu _1(A)^2$ the quantum form $Q_q$ is given by
\begin{equation*}
Q_q(f)=\int f(x)f(y)d(\nu _1\times\nu _1)(x,y)=\sqbrac{\int f(x)d\nu _1(x)}^2
\end{equation*}

Although the general quantum form $Q_q$ may be useful for certain applications, we do not believe that $Q_q(f)$ is a good candidate for the quantum integral of $f$. One reason is that it does not generalize the Lebesgue integral and hence, does not capture some of the properties that we think an integral should have. More precisely, if $\mu$ happens to be a measure, then $Q_q(f)$ is not the Lebesgue integral $\int fd\mu$.

\begin{thm}       
\label{thm36}
If $\mu$ is a measure, then the corresponding quantum form $Q_q$ satisfies $Q_q(f)=\int f^2d\mu$.
\end{thm}
\begin{proof}
If $\mu\colon\ascript\to\real^+$ is a measure, then the corresponding signed measure on
$\lambda\colon\ascript\times\ascript\to\real$ satisfies $\lambda (A\times B)=\mu (A\cap B)$. Let $f\in L_2(X)$ be a simple function given by
\begin{equation*}
f=\sum _{i=1}^nc_i\chi _{A_i}
\end{equation*}
where $c_i\in\real$ satisfy $c_i\ne c_j$, $i\ne j$ and $A_i\cap A_j=\emptyset$, $i\ne j$. We then have
\begin{align*}
Q_q(f)&=\int\sum c_i\chi _{A_i}(x)\sum c_j\chi _{A_j}(y)d\nu (x,y)\\
  &=\sum _{i,j}c_ic_j\int\chi _{A_i}(x)\chi _{A_j}(y)d\nu (x,y)\\
  &=\sum _{i,j}c_ic_j\int\chi _{A_i\times A_j}(x,y)d\nu (x,y)\\
  &=\sum _{i,j}c_ic_j\nu (A_i\times A_j)=\sum _{i,j}c_ic_j\mu (A_i\cap A_j)\\
  &=\sum _ic_i^2\mu (A_i)=\int f^2d\mu
\end{align*}
Since a nonnegative integrable function $f$ is a limit of an increasing sequence of simple functions the result holds for $f$. Since any integrable function $f$ can be written $f=f_1-f_2$ where $f_1,f_2\ge 0$ the result holds in general.
\end{proof}

\section{Quantum Integrals} 
Let $(X,\ascript ,\mu )$ be a $q$-measure space. We first discuss how one should \textit{not} define a quantum integral. The naive way is to follow Lebesgue. If $f$ is a simple measurable function (that is, $f$ has finitely many values), then $f$ has a unique canonical representation $f=\sum _{i=1}^nc_i\chi _{A_i}$ where $c_i\ne c_j$,
$A_i\cap A_j\ne\emptyset$, $i\ne j$, $A_i\in\ascript$. We then define the naive integral
\begin{equation*}
N\int fd\mu =\sum _{i=1}^nc_i\mu (A_i)
\end{equation*}
One problem is that $N\int fd\mu$ is ambiguous. Unlike the Lebesgue integral, if we represent $f$ in a noncanonical way $f=\sum d_i\chi _{B_i}$, then in general $N\int fd\mu\ne\sum d_i\mu (B_i)$. Another problem is that the usual limit theorems do not hold so there seems to be no way to extend this naive integral to arbitrary measurable functions. For instance, in Example~2, $N\int 1d\mu =1$. If we define the functions
\begin{equation*}
f_n=\chi _{\brac{x_1,x_2}}+\paren{1-\tfrac{1}{n}}\chi _{\brac{x_3}}
\end{equation*}
$n=1,2,\ldots$, then $f_n$ is an increasing sequence converging to $1$. However,
\begin{equation*}
N\int f_nd\mu =\mu\paren{\brac{x_1,x_2}}+\paren{1-\tfrac{1}{n}}\mu (x_3)=2-\tfrac{1}{n}
\end{equation*}
Hence,
\begin{equation*}
\lim _{n\to\infty}N\int f_nd\mu =2\ne 1=N\int 1d\mu
\end{equation*}

We now overcome the difficulties considered in the previous paragraph. Let $f\colon X\to\real$ be a measurable function and define the function $g\colon\sqpar{0,\infty}\to\real ^+$ by
\begin{equation*}
g(\lambda )=\mu\sqbrac{f^{-1}(\lambda ,\infty )}
\end{equation*}
If $f$ is simple, then $g$ is a step function and hence, $g$ is Lebesgue measurable. For arbitrary measurable $f$, there exists an increasing sequence of simple functions $f_i$ converging to $f$. Now
$f_i^{-1}(\lambda ,\infty )\subseteq f_{i+1}^{-1}(\lambda ,\infty )$ and
\begin{equation*}
f^{-1}(\lambda ,\infty )=\bigcup _{i=1}^\infty f_i^{-1}(\lambda ,\infty )
\end{equation*}
Defining $g_i(\lambda )=\mu\sqbrac{f_i^{-1}(\lambda ,\infty )}$, since $\mu$ is continuous, we have
\begin{align*}
g(\lambda )&=\mu\sqbrac{f^{-1}(\lambda ,\infty )}=\mu\sqbrac{\bigcup _{i=1}^\infty f_i^{-1}(\lambda ,\infty )}
  =\lim _{i\to\infty}\mu\sqbrac{f_i^{-1}(\lambda ,\infty )}\\
  &=\lim _{i\to\infty}g_i(\lambda )
\end{align*}
Since $g$ is a limit of measurable functions we conclude that $g$ is Lebesgue measurable. In a similar way, the function $h\colon\sqpar{0,\infty}\to\real ^+$ given by
\begin{equation*}
h(\lambda )=\mu\sqbrac{f^{-1}(-\infty ,-\lambda )}
\end{equation*}
is Lebesgue measurable. Denoting Lebesgue measure on $\real$ by $d\lambda$, we conclude that
$\int g(\lambda )d\lambda$ and $\int h(\lambda )d\lambda$ are defined. If both of these integrals are finite we say that $f$ is $\mu$-\textit{integrable} and we define
\begin{equation*}
\int fd\mu =\int g(\lambda )d\lambda -\int h(\lambda )d\lambda
\end{equation*}
In summary, the $q$-\textit{integral} of $f$ is
\begin{equation}
\label{eq41}              
\int fd\mu =\int _0^\infty\mu\sqbrac{f^{-1}(\lambda ,\infty )}d\lambda
  -\int _0^\infty\mu\sqbrac{f^{-1}(-\infty , -\lambda )}d\lambda
\end{equation}

Any measurable function $f\colon X\to\real$ has a unique representation $f=f_1-f_2$ where $f_1,f_2\ge 0$ are measurable and $f_1f_2=0$. In fact, $f_1=\max (f,0)$ and $f_2=-\min (f,0)$. We then have
\begin{align}
\label{eq42}              
\int fd\mu&=\int _0^\infty\mu\sqbrac{f_1^{-1}(\lambda ,\infty )}d\lambda
  -\int _0^\infty\mu\sqbrac{(-f_2)^{-1}(-\infty ,-\lambda )}d\lambda\notag\\
  &=\int f_1d\mu -\int f_2d\mu
\end{align}
Because of \eqref{eq42}, we can usually study the properties of the $q$-integral by considering nonnegative functions. We now derive some of the basic properties of the $q$-integral.

\begin{lem}       
\label{lem41}
Assume that $f\colon X\to\real$ is $\mu$-integrable.
{\rm (a)}\enspace If $f\ge 0$ then $\int fd\mu\ge 0$.
{\rm (b)}\enspace $\int cfd\mu =c\int fd\mu$ for all $c\in\real$.
{\rm (c)}\enspace $\int (c+f)d\mu =c\mu (X)+\int fd\mu =\int cd\mu +\int fd\mu$.
\end{lem}
\begin{proof}
The proof of (a) is clear. (b)\enspace If $c=0$, then the proof is obvious. If $c>0$, then using \eqref{eq42} we have
\begin{equation*}
\int cfd\mu =\int cf_1d\mu -\int cf_2d\mu
\end{equation*}
Now we have
\begin{align*}
\int cf_1d\mu&=\int _0^\infty\mu\sqbrac{(cf_1)^{-1}(\lambda ,\infty )}d\lambda
  =\int _0^\infty\mu\sqbrac{f_1^{-1}\paren{\tfrac{\lambda}{c},\infty}}d\lambda\\
  &=c\int _0^\infty\mu\sqbrac{f_1^{-1}(\lambda ,\infty )}d\lambda =c\int f_1d\mu
\end{align*}
The proof is the same for $f_2$ so that (b) holds in this case. If $c<0$, the proof is similar.
(c)\enspace If $c>0$, then using \eqref{eq42} we have $c+f=(c+f_1)-f_2$. Hence,
\begin{align*}
\int (c+f)d\mu&=\int (c+f_1)d\mu -\int f_2d\mu\\
  &=\int _0^\infty\mu\paren{\brac{x\colon c+f_1(x)>\lambda}}d\lambda-\int f_2d\mu\\
  &=\int _0^\infty\mu\paren{\brac{x\colon f_1(x)>\lambda -c}}-\int f_2d\mu
\end{align*}
Making a change of variable $\lambda _1=\lambda -c$ gives
\begin{align*}
\int (c+f)d\mu&=\int _{-c}^\infty\mu\paren{\brac{x\colon f_1(x)>\lambda _1}}d\lambda _1-\int f_2d\mu\\
  &=\int _{-c}^0\mu\sqbrac{f_1^{-1}(\lambda _1,\infty )}d\lambda _1
  +\int _0^\infty\mu\sqbrac{f_1^{-1}(\lambda _1,\infty )}d\lambda _1-\int f_2d\mu\\
  &=\int _{-c}^0\mu (X)d\lambda _1+\int f_1d\mu -\int f_2d\mu =c\mu (X)+\int fd\mu
\end{align*}
If $c<0$, then $c+f=f_1-(f_2-c)$ and by the above, we have
\begin{align*}
\int (c+f)d\mu&=\int f_1d\mu -\int (f_2-c)d\mu =\int f_1d\mu -\sqbrac{\int f_2d\mu -c\mu (X)}\\
  &=c\mu (X)+\int fd\mu\qedhere
\end{align*}
\end{proof}

\begin{lem}       
\label{lem42}
If $f$ is a simple measurable function
\begin{equation*}
f=\sum _{i=1}^n\alpha _i\chi _{A_i}
\end{equation*}
where $A_i\cap A_j=\emptyset$ for $i\ne j$ and $0<\alpha _1<\alpha _2<\cdots <\alpha _n$, then
\begin{align*}
\int fd\mu&=\alpha _1[\mu (A_1\cupdot A_2)+\cdots +\mu (A_1\cupdot A_n)\\
  &\quad -(n-2)\mu (A_1)-\mu (A_2)-\cdots -\mu (A_n)]\\
  &\quad +\alpha _2[\mu (A_2\cupdot A_3)+\cdots +\mu (A_2\cupdot A_n)\\
  &\quad -(n-3)\mu (A_2)-\mu (A_3)-\cdots -\mu (A_n)]\\
  &\quad\ \vdots\\
  &\quad +\alpha _{n-1}\sqbrac{\mu (A_{n-1}\cupdot A_n)-\mu (A_n)}+\alpha _n\mu (A_n)\\
\end{align*}
\end{lem}
\begin{proof}
Let $\alpha _0=0$ and for $i=0,1,\ldots, n$, let
\begin{equation*}
P(\alpha _i)=\mu\sqbrac{f^{-1}(\alpha _i,\infty )}=\mu (A_{i+1}\cupdot\cdots\cupdot A_n)
\end{equation*}
We then have by Theorem~\ref{thm22} that
\begin{align*}
\int&fd\mu
  =P(\alpha _0)\alpha _1+P(\alpha _1)(\alpha _2-\alpha _1)+\cdots +P(\alpha _{n-1})(\alpha _n-\alpha _{n-1})\\
  &=\alpha _1\sqbrac{\mu\paren{\bigcupdotin A_i}}+(\alpha _2-\alpha _1)\sqbrac{\mu\paren{\bigcupdotitwon A_i}}\\
  &\quad +\cdots +(\alpha _n-\alpha _{n-1}\sqbrac{\mu (A_n)}\\
  &=\alpha _1\sqbrac{\mu (A_1\cupdot A_2)+\cdots +\mu (A_{n-1}\cupdot A_n)-(n-2)\sum _{i=1}^n\mu (A_i)}\\
  &\quad +(\alpha _2-\alpha _1)\sqbrac{\mu (A_2\cupdot A_3)+\cdots +\mu (A_{n-1}\cupdot A_n)
  -(n-3)\sum _{i=2}^n\mu (A_i)}\\
  &\quad\ \vdots\\
  &\quad +(\alpha _{n-1}-\alpha _{n-2})\sqbrac{\mu A_{n-1}\cupdot A_n)}+(\alpha _n-\alpha _{n-1})\mu (A_n)
\end{align*}
and this reduces to the given expression.
\end{proof}

It follows from Lemma~\ref{lem42} that we have
\begin{align*}
\int&\alpha _1\chi _{A_1}d\mu =\alpha _1\mu (A_1)\\
  \int&(\alpha _1\chi _{A_1}+\alpha _2\chi _{A_2})d\mu =\alpha _1\sqbrac{\mu (A_1\cupdot A_2)-\mu (A_2)}
  +\alpha _2\mu (A_2)\\
  \int&(\alpha _1\chi _{A_1}+\alpha _2\chi _{A_2}+\alpha _3\chi _{A_3})d\mu\\
 & =\alpha _1\sqbrac{\mu (A_1\cupdot A_2)+\mu (A_1\cupdot A_3)-\mu (A_1)-\mu (A_2)-\mu (A_3)}\\
  &\quad +\alpha _2\sqbrac{\mu (A_2\cupdot A_3)-\mu (A_3)}+\alpha _3\mu (A_3)
\end{align*}
In general, the $q$-integral is not linear. For example, if $\mu (A\cupdot B)\ne\mu (A)+\mu (B)$, then
\begin{align*}
\int (\chi _A+\chi _B)d\mu&=\int\chi _{A\subcupdot B}d\mu=\mu (A\cupdot B)\ne\mu (A)+\mu (B)\\
   &=\int\chi _Ad\mu +\int\chi _Bd\mu
\end{align*}
In fact, we have the following result.

\begin{lem}       
\label{lem43}
If $A,B\in\ascript$ are arbitrary, then
\begin{equation*}
\int (\chi _A+\chi _B)d\mu =\int\chi _Ad\mu +\int\chi _Bd\mu
\end{equation*}
if and only if $A\mu B$
\end{lem}
\begin{proof}
Since $\chi _A+\chi _B=2\chi _{A\cap B}+\chi _{A\Delta B}$ we have by Lemma~\ref{lem42} that 
\begin{align*}
\int (\chi _A+\chi _B)d\mu&=\int (2\chi _{A\cap B}+\chi _{A\Delta B})d\mu\\
  &=\sqbrac{\mu (A\cup B)-\mu (A\cap B)}+2\mu (A\cap B)\\
  &=\mu (A\cup B)+\mu (A\cap B)
\end{align*}
This expression coincides with
\begin{equation*}
\int\chi _Ad\mu +\int\chi _Bd\mu =\mu (A)+\mu (B)
\end{equation*}
if and only if $A\mu B$.
\end{proof}

We conclude from Lemma~\ref{lem42} that if $\mu$ happens to be a (grade-1) measure, then
\begin{equation}
\label{eq43}       
\int fd\mu =\sum _{i=1}^n\alpha _i\mu (A_i)
\end{equation}
for a nonnegative simple function $f$. By Lemma~\ref{lem41}(b), \eqref{eq43} also holds for nonpositive simple functions and by \eqref{eq42} we have that \eqref{eq43} holds for any simple function. Thus, the $q$-integral generalizes the Lebesgue integral. The next result shows that a type of grade-2 additivity holds for the $q$-integral.

\begin{thm}       
\label{thm44}
If $f$, $g$ and $h$ are $\mu$-integrable functions with mutually disjoint support, then
\begin{equation*}
\int (f+g+h)d\mu =\!\!\int (f+g)d\mu +\int (f+h)d\mu +\!\int (g+h)d\mu -\!\int\!fd\mu -\!\int\!gd\mu -\!\int\!hd\mu
\end{equation*}
\end{thm}
\begin{proof}
First assume that $f,g,h\ge 0$. Since $\mu$ is grade-2 additive we have
\begin{align*}
\int &(f+g+h)d\mu\\
  &=\int _0^\infty\mu\paren{\brac{x\colon f(x)+g(x)+h(x)>\lambda}}d\lambda\\
  &=\int _0^\infty\mu\sqbrac{f^{-1}(\lambda ,\infty )\cupdot g^{-1}(\lambda ,\infty )\cupdot
    h^{-1}(\lambda ,\infty )}d\lambda\\
  &=\int _0^\infty\mu\sqbrac{f^{-1}(\lambda ,\infty )\cupdot g^{-1}(\lambda ,\infty )}d\lambda\\
  &\quad +\int _0^\infty\!\!\mu\sqbrac{f^{-1}(\lambda ,\infty )\cupdot h^{-1}(\lambda ,\infty )}d\lambda
    +\int _0^\infty\!\!\mu\sqbrac{g^{-1}(\lambda ,\infty )\cupdot h^{-1}(\lambda ,\infty )}d\lambda\\
  &\quad -\int _0^\infty\mu\sqbrac{f^{-1}(\lambda ,\infty )}d\lambda
    -\int _0^\infty\mu\sqbrac{g^{-1}(\lambda ,\infty )}d\lambda -\int\mu\sqbrac{h^{-1}(\lambda ,\infty )}d\lambda\\
  &=\int _0^\infty\mu\paren{\brac{x\colon f(x)+g(x)>\lambda}}d\lambda
  +\int _0^\infty\mu\paren{\brac{x\colon f(x)+h(x)>\lambda}}d\lambda\\
  &\quad +\int _0^\infty\mu\paren{\brac{x\colon g(x)+h(x)>\lambda}}d\lambda
  -\int  fd\mu -\int gd\mu -\int hd\mu\\
  &=\int (f+g)d\mu +\int (f+h)d\mu +\int (g+h)d\mu -\int fd\mu -\int gd\mu -\int hd\mu
\end{align*}
In a similar way the result holds if $f,g,h\le 0$. In the notation of \eqref{eq42}, since $(f+g+h)_1=f_1+g_1+h_1$ and
$(f+g+h)_2=f_2+g_2+h_2$ we have for general $f$, $g$ and $h$ that
\begin{align*}
\int&(f+g+h)d\mu\\
  &=\int (f_1+g_1+h_1)d\mu -\int (f_2+g_2+h_2)d\mu\\
  &=\int (f+g)_1d\mu -\int (f+g)_2d\mu +\int (f+h)_1d\mu -\int (f+h)_2d\mu\\
  &\quad +\int (g+h)_1d\mu -\int (g+h)_2d\mu -\paren{\int f_1d\mu -\int f_2d\mu}\\
  &\quad -\paren{\int g_1d\mu -\int g_2d\mu}-\paren{\int h_1d\mu -\int h_2d\mu}\\
  &=\int (f+g)d\mu +\int (f+h)d\mu +\int (g+h)d\mu -\!\!\int\!\!fd\mu -\!\!\int\!\!gd\mu -\!\!\int\!\!hd\mu\quad\qedhere
\end{align*}
\end{proof}

As with Lebesgue integrals, for a $q$-integral with $A\in\ascript$ we define
\begin{equation*}
\int _Afd\mu =\int f\chi _Ad\mu
\end{equation*}

\begin{cor}       
\label{cor45}
If $A,B,C\in\ascript$ are mutually disjoint, then
\begin{equation*}
\int _{A\subcupdot B\subcupdot C}fd\mu =\int _{A\subcupdot B}fd\mu +\int _{A\subcupdot C}fd\mu
  +\int_{B\subcupdot C}fd\mu -\int _Afd\mu -\int _Bfd\mu -\int _Cfd\mu
\end{equation*}
\end{cor}

By induction, Theorem~\ref{thm44} extends to $n$ integrable functions $f_1,\ldots ,f_n$ with mutually disjoint support:
\begin{equation*}
\int \sum _{i=1}^nf_id\mu =\sum _{i<j=1}^n\int (f_i+f_j)d\mu -(n-2)\sum _{i=1}^n\int f_id\mu
\end{equation*}
Theorem~\ref{thm44} does not hold for arbitrary integrable functions $f$, $g$ and $h$ so the strong type of grade-2 additivity considered in Section~3 does not hold for the $q$-integral. To illustrate this, consider Example~1 of a ``quantum coin.'' Let $A=\brac{x_1,x_2}$, $B=\brac{x_2,x_3}$, $C=\brac{x_3,x_4}$ and let $f=\chi _A$, $g=\chi _B$, $h=\chi _C$. Since
\begin{equation*}
f+g+h=\chi _{\brac{x_1,x_4}}+2\chi _{\brac{x_2,x_3}}
\end{equation*}
we have by Lemma~\ref{lem42} that
\begin{equation*}
\int (f+g+h)d\mu =\mu (X)-\mu (B)+2\mu (B)=1+\tfrac{1}{4}=\tfrac{5}{4}
\end{equation*}
However,
\begin{align*}
\int&(f+g)d\mu +\int (f+h)d\mu +\int (g+h)d\mu -\int fd\mu -\int gd\mu -\int hd\mu\\
  &=\int\paren{\chi _{\brac{x_1,x_3}}+2\chi _{\brac{x_2}}}d\mu +\int 1d\mu
  +\int\paren{\chi _{\brac{x_2,x_4}}+2\chi _{\brac{x_3}}}d\mu -\tfrac{3}{4}\\
  &=\mu\paren{\brac{x_1,x_2,x_3}}-\mu (x_2)+2\mu (x_2)+1+\mu\paren{\brac{x_2,x_3,x_4}}\\
  &\quad -\mu (x_3)+2\mu (x_3)-\tfrac{3}{4}\\
  &=\tfrac{3}{2}
\end{align*}
These two expressions do not coincide. To further illustrate the $q$-integral, let $f$ be the number of heads given by $f(x_1)=2$, $f(x_2)=f(x_3)=1$, $f(x_4)=0$. We can then write
\begin{equation*}
f=\chi _{\brac{x_2,x_3}}+2\chi _{\brac{x_1}}
\end{equation*}
Hence,
\begin{equation*}
\int fd\mu =\mu\paren{\brac{x_1,x_2,x_3}}-\mu (x_1)+2\mu (x_1)=\tfrac{9}{16}+\tfrac{1}{16}=\tfrac{5}{8}
\end{equation*}
In comparison, the naive integral is
\begin{equation*}
N\int fd\mu =\mu\paren{\brac{x_2,x_3}}+2\mu (x_1)=\tfrac{3}{8}
\end{equation*}
Finally, we illustrate the $q$-integral for the $q$-measure $\mu$ in Example~4. Let $f(x)=x$ for $x\in\sqbrac{0,1}$. Of course, the Lebesgue (or Reimann) integral gives $\int _0^1f(x)dx=1/2$. However, the $q$-integral becomes:
\begin{align*}
\int fd\mu&=\int _0^{1/4}\paren{\tfrac{1}{2}+\lambda}d\lambda +\int _{1/4}^1(1-\lambda )d\lambda\\
  &=\sqbrac{\tfrac{1}{2}\lambda +\tfrac{1}{2}\lambda ^2}_0^{1/4}+\sqbrac{\lambda -\tfrac{1}{2}\lambda ^2}_{1/4}^1
  =\tfrac{7}{16}
\end{align*}

\section{Convergence Theorem} 
Let $(X,\ascript ,\mu )$ be a $q$-measure space. For measurable functions $f$, $g$ on $X$ we say that $g$
$\mu$-\textit{dominates} $f$ if
\begin{equation*}
\mu\sqbrac{f^{-1}(\lambda ,\infty )}\le\mu\sqbrac{g^{-1}(\lambda ,\infty )}
\end{equation*}
for all $\lambda\in\real$. We now show that this is a weaker concept than the usual domination for (grade-1) measures.

\begin{lem}       
\label{lem51}
If $\nu$ is a measure, the $f\le g\hbox{ a.e.\,}\sqbrac{\nu}$ implies that $g$ $\nu$-dominates $f$.
\end{lem}
\begin{proof}
Since $f\le g\hbox{ a.e.\,}\sqbrac{\nu}$, if $A=\brac{x\colon g(x)<f(x)}$ we have that $\nu (A)=0$. Since
\begin{equation*}
f^{-1}(\lambda ,\infty )\cap A'\subseteq g^{-1}(\lambda ,\infty )\cap A'
\end{equation*}
we conclude that
\begin{align*}
\nu\sqbrac{f^{-1}(\lambda ,\infty )}&=\nu\sqbrac{f^{-1}(\lambda ,\infty )\cap A'}
  +\mu\sqbrac{f^{-1}(\lambda ,\infty )\cap A}\\
  &\le\nu\sqbrac{g^{-1}(\lambda ,\infty )\cap A'}=\nu\sqbrac{g^{-1}(\lambda ,\infty )}\qedhere
\end{align*}
\end{proof}

The converse of Lemma~\ref{lem51} does not hold. For example, let $X=\sqbrac{0,1}$ and let $\nu$ be Lebesgue measure on $X$. If
\begin{equation*}
f=\tfrac{1}{2}\chi _{\sqbrac{1/2,1}},\quad g=\chi _{\sqbrac{0,1/2}}
\end{equation*}
Then $g$ $\nu$-dominates $f$ but $f\not\le g\hbox{ a.e.\,}\sqbrac{\nu}$. We call the next result the $q$-dominated, monotone convergence theorem.

\begin{thm}       
\label{thm52}
If $f_i\ge 0$ is an increasing sequence of measurable functions on $X$ that converge to $f$ and there exists an integrable function $g$ such that $g$ $\mu$-dominates $f_i$ for all $i$, then
\begin{equation*}
\lim _{i\to\infty}\int f_id\mu =\int fd\mu
\end{equation*}
\end{thm}
\begin{proof}
Define the measurable functions $u(\lambda )=\mu\sqbrac{f^{-1}(\lambda ,\infty )}$ and
$u_i(\lambda )=\mu\sqbrac{f_i^{-1}(\lambda ,\infty )}$, $\lambda\ge 0$. Since
$f_i^{-1}(\lambda ,\infty )\subseteq f_{i+1}^{-1}(\lambda ,\infty )$ and
$f^{-1}(\lambda ,\infty )=\cup f_i^{-1}(\lambda ,\infty )$ we have by the continuity of $\mu$ that
\begin{equation*}
u(\lambda )=\mu\sqbrac{\cup f_i^{-1}(\lambda ,\infty )}=\lim _{i\to\infty}\mu\sqbrac{f_i^{-1}(\lambda ,\infty )}
  =\lim _{i\to\infty}u_i(\lambda )
\end{equation*}
Since $g$ $\mu$-dominates $f_i$, letting $v(\lambda )=\mu\sqbrac{g^{-1}(\lambda ,\infty )}$ we have that $u_i(\lambda )\le v(\lambda )$ for all $\lambda$ and
\begin{equation*}
\int _0^\infty v(\lambda )d\lambda =\int gd\mu <\infty
\end{equation*}
By the dominated convergence theorem we conclude that
\begin{equation*}
\lim _{i\to\infty}\int f_id\mu =\lim _{i\to\infty}\int _0^\infty u_i(\lambda )d\lambda
  =\int _0^\infty u(\lambda )d\lambda =\int fd\mu\qedhere
\end{equation*}
\end{proof}

Let $f\colon X\to\real ^+$ be measurable and suppose there exists a Lebesgue integrable function
$g\colon\real\to\real^+$ such that for every $A\in\ascript$ we have
\begin{equation*}
\mu\paren{\brac{x\in A\colon f(x)>\lambda}}\le g(\lambda )
\end{equation*}
for all $\lambda\in\real$. Defining $\mu _1(A)=\int _Afd\mu$ we conclude that
\begin{equation*}
\mu _1(A)=\int\mu\paren{\brac{x\in A\colon f(x)>\lambda}}d\lambda\le\int g(\lambda )d\lambda <\infty
\end{equation*}
for all $A\in\ascript$. It follows from Corollary~\ref{cor45} that $\mu _1$ is grade-2 additive.

\begin{thm}       
\label{thm53}
{\rm (a)}\enspace $\mu _1$ is a $q$-measure on $\ascript$.
{\rm (b)}\enspace If $\mu$ is regular, then so is $\mu _1$.
{\rm (c)}\enspace If $\mu$ is completely regular, then so is $\mu _1$.
{\rm (d)}\enspace If $\mu$ is completely regular, then $\mu _1\ll\mu$ (that is, $\mu (A)=0$ implies that
$\mu _1(A)=0$).
\end{thm}
\begin{proof}
(a)\enspace Let $A_i\in\ascript$ be increasing and let $A=\cup A_i$. By the dominated convergence theorem, we have
\begin{align*}
\mu _1(A)&=\int\mu\sqbrac{A\cap f^{-1}(\lambda ,\infty )}d\lambda
  =\int\mu\sqbrac{\cup\paren{A_i\cap f^{-1}(\lambda ,\infty )}}d\lambda\\
  &=\int\lim _{i\to\infty}\mu\sqbrac{A_i\cap f^{-1}(\lambda ,\infty )}d\lambda
    =\lim _{i\to\infty}\int\mu\sqbrac{A_i\cap f^{-1}(\lambda ,\infty )}d\lambda\\
    &=\lim\mu _1(A_i)
\end{align*}
If $A_i\in\ascript$ is decreasing, we obtain a similar result. Hence, $\mu _1$ is a grade-2 measure so $\mu _1$ is a
$q$-measure.
(b)\enspace Assume that $\mu$ is regular. If $\mu _1(A)=0$, then
\begin{equation*}
\int\mu\sqbrac{A\cap f^{-1}(\lambda ,\infty )}d\lambda =\mu _1(A)=0
\end{equation*}
Hence, $\mu\sqbrac{A\cap f^{-1}(\lambda ,\infty )}=0\hbox{ a.e.}$. We then have that
\begin{align*}
\mu _1(A\cupdot B)&=\int\mu\sqbrac{(A\cupdot B)\cap f^{-1}(\lambda ,\infty )}d\lambda\\
  &=\int\mu\sqbrac{\paren{A\cap f^{-1}(\lambda ,\infty )}\cupdot\paren{B\cap f^{-1}(\lambda ,\infty )}}d\lambda\\
  &=\int\mu\sqbrac{B\cap f^{-1}(\lambda ,\infty )}d\lambda =\mu _1(B)
\end{align*}
Moreover, if $\mu _1(A\cupdot B)=0$, then by the previous calculation we have that
\begin{equation*}
\mu\sqbrac{\paren{A\cap f^{-1}(\lambda ,\infty )}\cupdot\paren{B\cap f^{-1}(\lambda ,\infty )}}=0\hbox{ a.e.}
\end{equation*}
Since $\mu$ is regular,
\begin{equation*}
\mu\sqbrac{A\cap f^{-1}(\lambda ,\infty )}=\mu\sqbrac{B\cap f^{-1}(\lambda ,\infty )}\hbox{ a.e.}
\end{equation*}
It follows that $\mu _1(A)=\mu _1(B)$. Hence, $\mu _1$ is regular.
(c)\enspace Assume that $\mu$ is completely regular. Then $\mu _1$ is regular by (b). If $\mu _1(A)=0$ and
$B\subseteq A$ with $B\in\ascript$, then as before $\mu\sqbrac{A\cap f^{-1}(\lambda ,\infty )}=0\hbox{ a.e.}$ so
$\mu\sqbrac{B\cap f^{-1}(\lambda ,\infty )}=0\hbox{ a.e.}$ It follows that $\mu _1(B)=0$. Hence, $\mu _1$ is completely regular.
(d)\enspace Again, assume that $\mu$ is completely regular. If $\mu (A)=0$, then
$\mu\sqbrac{A\cap f^{-1}(\lambda ,\infty )}=0$ for all $\lambda\in\real$. Hence, $\mu _1(A)=0$.
\end{proof}

Theorem~\ref{thm53}(d) suggests a quantum Radon-Nikodym theorem for completely regular $q$-measures. Unfortunately the next counterexample shows that no such theorem holds even when $X$ is finite. As in Example~2, let $X=\brac{x_1,x_2,x_3}$ and let $\mu$ be the completely regular $q$-measure given by
$\mu (\emptyset )=\mu (x_1)=0$ and $\mu (A)=1$ for all other $A\in\pscript (X)$. Let $\nu$ be the measure on
$\pscript (X)$ given by $\nu (x_1)=0$, $\nu (x_2)=\nu (x_3)=1$ so that
\begin{equation*}
\nu\paren{\brac{x_2,x_3}}=\nu (X)=2
\end{equation*}
and
\begin{equation*}
\nu\paren{\brac{x_1,x_2}}=\nu\paren{\brac{x_1,x_3}}=1
\end{equation*}
Then $\nu\ll\mu$. Suppose there exists a function $f\ge 0$ such that $\nu (A)=\int _Afd\mu$ for all $A\in\pscript (X)$. Then
\begin{equation*}
f(x_2)=f(x_2)\mu (x_2)=\int _{\brac{x_2}}fd\mu =\nu (x_2)=1
\end{equation*}
and similarly, $f(x_3)=1$. Hence,
\begin{equation*}
2=\nu\paren{\brac{x_2,x_3}}=\int _{\brac{x_2,x_3}}fd\mu =\mu\paren{\brac{x_2,x_3}}=1
\end{equation*}
which is a contradiction.

\section{Quantum Lebesgue Measure} 
This section explores a particularly interesting example of a $q$-measure and its corresponding $q$-integral. Let
$X=\sqbrac{0,1}$ and let $\nu$ be Lebesgue measure on $\bscript (X)$. Define the $q$-measure $\mu$ on
$\bscript (X)$ by $\mu (A)=\nu (A)^2$. We call $\mu$ $q$-\textit{Lebesgue measure}. In the sequel, $y$ will denote a fixed element of $X$. The next result gives the $q$-Lebesgue integral for the general monomial $f(x)=x^n$.

\begin{thm}       
\label{thm61}
For $n=0,1,\ldots$, we have
\begin{equation*}
\int _{\sqbrac{0,y}}x^nd\mu (x)=\frac{2}{(n+1)(n+2)}\,y^{n+2}
\end{equation*}
\end{thm}
\begin{proof}
Since $f(x)=x^n$ is invertible and increasing we have
\begin{align*}
\int _{\sqbrac{0,y}}x^nd\mu (x)
  &=\int _0^\infty\nu\paren{\brac{x\colon x^n\chi _{\sqbrac{0,y}}(x)>\lambda}}^2d\lambda\\
  &=\int _0^{y^n}\nu\paren{\brac{x\colon x\chi _{\sqbrac{0,y}}(x)>\lambda ^{1/n}}}^2d\lambda\\
  &=\int _0^{y^n}(y-\lambda ^{1/n})^2d\lambda =\int _0^{y^n}(y^2-2y\lambda ^{1/n}+\lambda ^{2/n})d\lambda\\
  \noalign{\medskip}
  &=\sqbrac{y^2\lambda-\frac{2y^{1+1/n}}{1+1/n}+\frac{\lambda ^{1+2/n}}{1+2/n}}_0^{y^n}\\
  \noalign{\medskip}
  &=\paren{1-\frac{2n}{n+1}+\frac{n}{n+2}}\,y^{n+2}=\frac{2}{(n+1)(n+2)}\,y^{n+2}\qedhere
\end{align*}
\end{proof}

For example, it follows from Theorem~\ref{thm61} that
\begin{align*}
\int _{\sqbrac{0,y}}1d\mu (x)=y^2\\
\int _{\sqbrac{0,y}}xd\mu (x)=\tfrac{1}{3}\,y^3\\
\int _{\sqbrac{0,y}}x^2d\mu (x)=\tfrac{1}{6}\,y^4
\end{align*}
We next compute the $q$-integral of $e^x$.
\begin{align*}
\int _0^ye^xd\mu (x)&=\int _0^\infty\nu\paren{\brac{x\colon e^x\chi _{\sqbrac{0,y}}(x)>\lambda}}^2d\lambda\\
  &=\int _0^{e^y}\nu\paren{\brac{x\colon x\chi _{\sqbrac{0,y}}(x)>\ln\lambda}}^2d\lambda\\
  &=\int _0^1y^2d\lambda +\int _1^{e^y}(y-\ln\lambda )^2d\lambda\\
  &=y^2+\sqbrac{y^2\lambda -2y(\lambda\ln\lambda -\lambda )
    +\lambda (\ln\lambda )^2-2\lambda\ln\lambda +2\lambda}_1^{e^y}\\
  &=2(e^y-y-1)
\end{align*}
Although the $q$-Lebesgue integral is nonlinear we have the surprising result that
\begin{equation*}
\int _{\sqbrac{0,y}}(x^2+x)d\mu (x)=\int _{\sqbrac{0,y}}x^2d\mu (x)+\int _{\sqbrac{0,y}}xd\mu (x)
\end{equation*}
Indeed, letting $f(x)=x^2+x$ we find that
\begin{equation*}
f^{-1}(x)=-\tfrac{1}{2}+\tfrac{1}{2}\,\sqrt{1+4x\,}
\end{equation*}
for $x\in\sqbrac{0,1}$. We then obtain
\begin{align*}
\int _{\sqbrac{0,y}}(x^2+x)d\mu (x)
  &=\int _0^\infty\nu\paren{\brac{x\colon (x^2+x)\chi _{\sqbrac{0,y}}(x)>\lambda}}^2d\mu\\
  &=\int _0^{y^2+y}\nu\paren{\brac{x\colon x\chi _{\sqbrac{0,y}}(x)>f^{-1}(\lambda )}}^2d\mu\\
  &=\int _0^{y^2+y}\paren{y+\tfrac{1}{2}-\tfrac{1}{2}\,\sqrt{1+4\lambda\,}}^2d\lambda\\
  &=\int _0^{y^2+y}\sqbrac{\paren{y+\tfrac{1}{2}}^2-\paren{y+\tfrac{1}{2}}\,\sqrt{1+4\lambda\,}
    +\tfrac{1}{4}\,(1+4\lambda )}d\lambda\\
  &=\sqbrac{\paren{y+\tfrac{1}{2}}^2\lambda -\tfrac{1}{6}\,\paren{y+\tfrac{1}{2}}(1+4\lambda )^{3/2}
    +\tfrac{1}{4}\lambda +\tfrac{1}{2}\lambda ^2}_0^{y^2+y}\\
  &=\tfrac{1}{6}\,y^4+\tfrac{1}{3}\,y^3=\int _{\sqbrac{0,y}}x^2d\mu (x)+\int _{\sqbrac{0,y}}xd\mu (x)
\end{align*}

We call the next result the quantum fundamental theorem of calculus.
\begin{thm}       
\label{thm62}
If $f$ is twice differentiable and monotone on $(0,y)$, then
\begin{equation*}
\frac{1}{2}\,\frac{d^2}{dy^2}\int _{\sqbrac{0,y}}f(x)d\mu (x)=f(y)
\end{equation*}
\end{thm}
\begin{proof}
We assume that $f$ is increasing. The proof for decreasing $f$ is similar. By the usual fundamental theorem of calculus, we have
\begin{align*}
\frac{1}{2}\,\frac{d^2}{dy^2}\int _{\sqbrac{0,y}}f(x)d\mu (x)
  &=\frac{1}{2}\,\frac{d^2}{dy^2}\int _0^\infty\nu\paren{\brac{x\colon f(x)\chi _{\sqbrac{0,y}}(x)>\lambda}}^2d\lambda\\
  \noalign{\medskip}
  &=\frac{1}{2}\,\frac{d^2}{dy^2}\int _0^{f(y)}\nu
    \paren{\brac{x\colon x\chi _{\sqbrac{0,y}}(x)>f^{-1}(\lambda )}}^2d\lambda\\
   \noalign{\medskip}
  &=\frac{1}{2}\,\frac{d^2}{dy^2}\int _0^{f(y)}\sqbrac{y-f^{-1}(\lambda )}^2d\lambda\\
  \noalign{\medskip}
  &=\frac{1}{2}\,\frac{d^2}{dy^2}\int _0^{f(y)}\sqbrac{y^2-2yf^{-1}(\lambda )+f^{-1}(\lambda )^2}d\lambda\\
  \noalign{\medskip}
  &=\frac{1}{2}\,\frac{d^2}{dy^2}\sqbrac{y^2f(y)}-\frac{d^2}{dy^2}\sqbrac{y\int _0^{f(y)}f^{-1}(\lambda )d\lambda}\\
  \noalign{\medskip}
  &\quad +\frac{1}{2}\,\frac{d^2}{dy^2}\sqbrac{\int _0^{f(y)}f^{-1}(\lambda )^2d\lambda}\\
  \noalign{\medskip}
  &=\frac{1}{2}\,\frac{d}{dy}\sqbrac{y^2f'(y)+2yf(y)}\\
  &\quad -\frac{d}{dy}\sqbrac{y^2f'(y)+\int _0^{f(y)}f^{-1}(\lambda )d\lambda}
    +\frac{1}{2}\,\frac{d}{dy}\sqbrac{f'(y)y^2}\\
  &=yf'(y)+f(y)-f'(y)y=f(y)\qedhere
\end{align*}
\end{proof}


\begin{thebibliography}{99}
\bibitem{gud1}S.~Gudder, Quantum measure theory, preprint, Univ. of Denver.
\bibitem{gud2}S.~Gudder, Finite quantum measure spaces, \textit{Amer. Math. Monthly} (to appear).
\bibitem{sal02}R.~Salgado, Some identities for the quantum measure and its generalizations,
\textit{Mod. Phys.  Letts.~A} \textbf{17} (2002), 711--728.
\bibitem{scmslsw08}U.~Sinha, C.~Couteau, Z.~Mendendorp, I.~S\"ollner, R.~Laflamme, R.~Sorkin
and G.~Weihs, Testing Born's rule in quantum mechanics with a triple slit experiment,
arXiv: 0811.2068v1 [quant-ph] (2008).
\bibitem{sor94}R.~Sorkin, 
Quantum mechanics as quantum measure theory, \textit{Mod. Phys. Letts.~A} \textbf{9} (1994), 3119--3127.
\bibitem{sor07}R.~Sorkin, 
Quantum mechanics without the wave function, \textit{J.~Phys.~A} \textbf{40} (2007), 3207--3231.
\bibitem{sw08}S.~Surya and P.~Wallden, 
Quantum covers in quantum measure theory, arXiv: 0809.1951 [quant-ph] (2008).
\end{thebibliography}
\end{document}